\documentclass[longauth]{aa} 
\usepackage{graphicx}
\usepackage{amsmath}
\usepackage{amssymb}
\usepackage{float}
\usepackage{natbib}
\bibpunct{(}{)}{;}{a}{}{,}

\usepackage{txfonts}
\begin{document}

\title{Morphological and spectral properties of the W51 region measured with the MAGIC telescopes}

\author{
 J.~Aleksi\'c\inst{1} \and
 E.~A.~Alvarez\inst{2} \and
 L.~A.~Antonelli\inst{3} \and
 P.~Antoranz\inst{4} \and
 M.~Asensio\inst{2} \and
 M.~Backes\inst{5} \and
 U.~Barres de Almeida\inst{6} \and
 J.~A.~Barrio\inst{2} \and
 D.~Bastieri\inst{7} \and
 J.~Becerra Gonz\'alez\inst{8,}\inst{9}
% ...
}
\institute { IFAE, Edifici Cn., Campus UAB, E-08193 Bellaterra, Spain
 \and Universidad Complutense, E-28040 Madrid, Spain
 \and INAF National Institute for Astrophysics, I-00136 Rome, Italy
 \and Universit\`a  di Siena, and INFN Pisa, I-53100 Siena, Italy
 \and Technische Universit\"at Dortmund, D-44221 Dortmund, Germany
 \and Max-Planck-Institut f\"ur Physik, D-80805 M\"unchen, Germany
 \and Universit\`a di Padova and INFN, I-35131 Padova, Italy
 \and Inst. de Astrof\'{\i}sica de Canarias, E-38200 La Laguna, Tenerife, Spain
 \and Depto. de Astrof\'{\i}sica, Universidad de La Laguna, E-38206 La Laguna, Spain
% ...
}

\date{Received: 19 January 2012 / Accepted: 03 February 2012}

\offprints{J.~Krause (julkrau@googlemail.com), I.~Reichardt (ignasi@ifae.cat) and E.~Carmona (emilianocarm@googlemail.com)}


  \abstract
   {The W51 complex hosts the supernova remnant W51C which is known to interact with the molecular clouds in the star forming region W51B. In addition,
a possible pulsar wind nebula CXO J192318.5+140305 was found likely associated with the supernova remnant.
Gamma-ray emission from this region was discovered by \textit{Fermi}/LAT (between 0.2 and 50 GeV) and H.E.S.S. ($>$1 TeV).
The spatial distribution of the events could not be used to pinpoint the location of the emission among the pulsar wind nebula,
the supernova remnant shell and/or the molecular cloud. However, the modeling of the spectral energy distribution presented by the \textit{Fermi}/LAT 
collaboration suggests a hadronic emission mechanism. The possibility that the gamma-ray emission from such an object is of hadronic origin can contribute to solvingthe long-standing problem of the contribution to galactic cosmic rays by supernova remnants.}
   {Our aim is to determine the morphology of the very-high-energy gamma-ray emission of W51 and measure its spectral properties.}
   {We performed observations of the W51 complex with the MAGIC telescopes for more than 50 hours. The energy range accessible with MAGIC extends from $50$ GeV to several TeV,
allowing for the first spectral measurement at these energies. In addition, the good angular resolution in the medium (few hundred GeV) to high (above 1\,TeV)
energies allow us to perform morphological studies.
We look for underlying structures by means of detailed morphological studies.
Multi-wavelength data from
this source have been sampled to model the emission with both leptonic and hadronic processes.}
   {We detect an extended emission of very-high-energy gamma rays, with a significance of 11 standard deviations. We extend the spectrum from the
     highest \textit{Fermi}/LAT energies to $\sim$ 5\,TeV and find that it follows a single power law with an index of $2.58\pm0.07_{\tiny\mbox{stat}}\pm0.22_{\tiny\mbox{syst}}$.
The main part of the emission coincides with the shocked cloud region, while we find a feature extending towards the pulsar wind nebula. The possible contribution of the pulsar wind nebula,
assuming a point-like source, shows no dependence on energy and it is about 20\% of the overall emission. The broad band spectral energy distribution can be explained with a hadronic model that implies proton acceleration above 100\,TeV.
This result, together with the morphology of the source, tentatively suggests that we observe ongoing acceleration of ions in the interaction zone between supernova remnant and cloud.}
   {}
   \maketitle

\section{Introduction}
\label{intro}
W51 is a massive molecular complex located at the tangential point ($\textit{l} = 49^{\circ{}}$) of the Sagitarius arm of the Galaxy, at a distance of $\sim5.5$\,kpc
\citep{parallax}. As seen in radio continuum images, three main components are identified: the star-forming regions W51A and W51B and, attached to the south-eastern
boundary of W51B, the supernova remnant (SNR) W51C. The estimated age of this SNR is 30 kyrs \citep{KooRosat}. Evidence of interaction between W51C and W51B is provided by several observations. Most crucial
of them are the existence of two 1720 MHz OH masers \citep{maser} and the detection of about $10^{3}$ solar masses of atomic gas at a velocity shifted between 20 and
120\,km\,s$^{-1}$ with respect to its ambient medium \citep{kooHI}. The high-velocity atomic gas exhibits a counterpart in high density molecular gas clumps \citep{kooH2} sharing
the same location and velocity shift. Koo \& Moon showed that the shocked gas is displayed in a thin layer in the interface between the SNR shell, as delimited by the X-ray
image from \textit{ROSAT} and the \textit{unshocked} molecular gas. This can be taken as the existence of a J-type shock penetrating the dense gas in a particular region of W51B\citep{kooH2}, whereas in the location of the 1720 MHz OH masers the shock should be continuous (C-type). Moreover, recent measurements \citep{crionization} showed over-ionization of the gas in W51B in certain locations close to W51C coinciding with the shocked gas. They conclude this excess in ionization implies the existence of an intense flow of freshly accelerated cosmic rays (CRs) that, through proton-proton collisions, ionize the hydrogen in the adjacent cloud. However, $\sim0.2$ degrees South-East to the shocked gas region, a hard X-ray source CXO J192318.5+140305 is detected. This object was first resolved by \textit{ASCA} \citep{kooASCA} and later confirmed by \textit{Chandra}
\citep{kooChandra}. Its X-ray spectrum, together with its morphology, suggests that it is a possible pulsar wind nebula (PWN) associated with the SNR. Therefore, the presence of CXO J192318.5+140305 plays a role in the interpretation of the gamma-ray emission from the W51 region.
For these reasons, W51C represents an interesting case for the study of the acceleration of particles to very high energies (VHE) and their interaction with the interstellar medium.

An extended source of gamma rays was first detected by the H.E.S.S. telescopes with an integral flux above $1\,$TeV of about 3\% that of the Crab Nebula
\citep{W51Hess}. However, the presented morphological and spectral information was not enough to attribute the origin of the emission to any particular object in the field of view. Also, the Large Area Telescope (LAT) on board the \textit{Fermi} satellite detected an extended source between 200\,MeV and 50\,GeV coincident with the H.E.S.S. source
\citep{fermi}. Moreover, the reanalysis of the archival MILAGRO data after the release of the first \textit{Fermi} catalog revealed a 3.4$\sigma$ excess with median
energy of 10\,TeV coincident with the \textit{Fermi}/LAT source \citep{MILAGRO}.
At radio wavelengths, synchrotron radiation on ambient magnetic field explains the emission detected from W51C. 
At higher energies, there are several processes that yield emission of gamma rays: inverse Compton scattering of electrons on seed photons 
(cosmic microwave background, starlight), non-thermal bremsstrahlung of electrons on charged target, and decay of neutral pions created in flight from a 
proton-nucleon collision. The modeling done by \citep{fermi} of the spectral energy distribution (SED) of W51C disfavors leptonic models and suggests a 
hadronic origin for the emission.
For the hadronic channel, two main (non-exclusive) mechanisms are to be considered: molecular cloud illumination by cosmic rays that escaped the accelerating shock
 \citep{gabici,ohira} or emission from clouds that are being overtaken by the SNR blast wave \citep{uchiyama,fang}.
It is well known that a 10\% of the energy released by the supernova explosions in the Galaxy can account for the energy budget of the CR spectrum up to energies close to the \textit{knee} ($\sim10^{15}$ eV). 
Nevertheless, the evidence that SNRs can accelerate particles up to such high energies is still missing.
Since W51C is one of the most luminous Galactic sources at \textit{Fermi}/LAT energies, observation of gamma rays up to several TeV would have serious implications 
regarding the SNR contribution to the Galactic CRs: such an observation would show that SNRs are not only capable to provide a sufficient flux, 
but could also shed light on the question of the maximum energy of CR's still achievable in such a medium age SNR.

However, the object from which the gamma rays originate has not yet been identified within the W51 field, and the gamma-ray spectrum has so far been precisely measured only up to some tens of GeV.
In what follows, we report observations with the MAGIC telescopes, which will
help to address some of the remaining questions on the gamma-ray source in the W51 region, both regarding its precise location and the physical processes needed to explain the observations. In~\ref{sec:Obs} we describe the observations that we performed; in~\ref{sec:results} we show the observed morphology and spectral properties; and, finally, in~\ref{sec:Model} we apply a theoretical framework that can explain the detected gamma-ray emission.

\section{Observations}
\label{sec:Obs}
MAGIC consists of two $17\,\mathrm{m}$ diameter Imaging Atmospheric Cherenkov Telescopes
(IACT) located at the Roque de los Muchachos observatory, on La Palma island, Spain ($28^{\circ}46'\,$N,
$17^{\circ}53'\,$W), at the height of $2200$\,m a.s.l. The stereo observations provide a sensitivity\footnote{\small{Sensitivity is defined here as the minimal integral flux to reach $5\,\sigma$ excess in $50$\,h of observations, assuming a spectral index like that of the Crab Nebula.}} of $0.8\%$ of the Crab Nebula flux at energies $>300\,$GeV, see \cite{MAGICperformance}.
MAGIC has the lowest trigger threshold of all operating IACTs, enabling it to observe gamma rays between 50\,GeV and several tens of TeV.
MAGIC observed W51 in 2010 and 2011. In the first period of observations between May 17 and August 19 2010 about 31 hours effective time remained after quality cuts.
Between May 3 and June 13 2011 additional 22 hours effective time of good quality data were taken, resulting in a total amount of 53 h effective dark time and covering a zenith angle range from 14 to 35 degrees.
The observations were carried out in the so-called wobble mode around the center of the \textit{Fermi}/LAT source W51C ($\mathrm{RA}=19.385
\,\mathrm{h}, \mathrm{DEC}=14.19^{\circ}$).
All data were taken in stereoscopic mode, recording only events which triggered both telescopes.
To minimize systematic effects in the exposure and to optimize the coverage for an unknown extension of the emission a total of six pointing positions ($n_{\mathrm{point}} = 6$),
were used. In all pointing positions the wobble distance (offset from the central position) was $0.4^{\circ}$, as it is regularly done in MAGIC observations.

The analysis of the data was performed using the MARS analysis framework \citep{mars} including the latest standard routines for stereoscopic analysis \citep{lombardi}.
After calibrating the signal and cleaning the images of the two telescopes individually, the two images of each stereo event are combined. The arrival direction is determined
from the combination of the individual telescope information. To suppress the background, a global variable dubbed
\textit{hadronness} is determined by using the so-called \textit{random forest} method \citep{RF}.
The energy of individual events is estimated using look-up tables generated from gamma-ray Monte-Carlo events.
For a detailed description of the complete analysis chain described above see \cite{MAGICperformance}.
The gamma-ray signal is estimated by comparing the spatial distribution of gamma-like events around the
assumed source position (ON region) with respect to those recorded in signal-free (OFF) regions.
The total signal of the source is evaluated using a cut on the squared angular distance between reconstructed gamma-ray direction and source position of $\theta^2<0.07$.
For each pointing position, the ON sample is compared to an OFF sample obtained from the combination of the
$n_{\mathrm{point}}-1$ OFF regions observed at the same focal plane coordinates but from the complementary  pointing positions.
Four of the pointing positions have an observation time of the order of $\sim12$ hours each. Therefore, three background samples per pointing can be averaged. The 
remaining two positions have an observation time around $2$ hours each and, in this case, the background was estimated from one sample only.
This method ensures a maximum usage of symmetrical OFF positions without introducing big scaling factors due to differences in the observation time.
The significance of the excess is determined from the combined $\theta^2$ distribution of all individual pointing positions using equation 17 in \cite{LiMa} 
(Li\&Ma significance hereafter).

\section{Results}
\label{sec:results}
\subsection{Detection}
\begin{figure}
   \centering
   \includegraphics[angle=0,width=88mm]{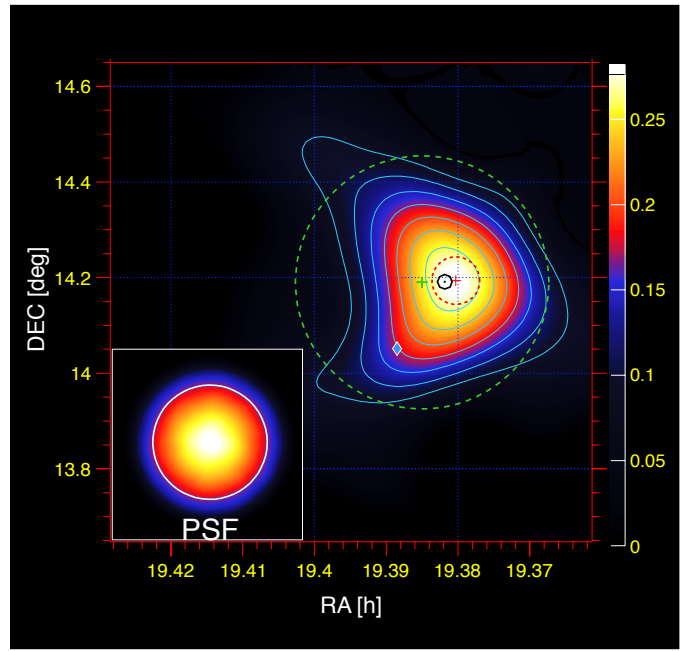}
   \caption{Relative flux (excess/background) map above $150$\,GeV around W51. Overlaid are contour levels from test statistics starting at 3 and increasing by one per contour.
The map was smoothed with a Gaussian kernel of 0.085\degr. The green cross represents the center of the observations, while the green dashed circle represents the
integration area. The black dot is the determined position of the centroid with the statistical uncertainties shown by the surrounding black ellipse. The region of
shocked atomic and molecular gas~\citep{kooH2, kooHI} is represented by the red dashed ellipse. The blue diamond shows the position of the possible PWN CXO J192318.5+140305. In the left lower corner the gaussian sigma of a point-like source (PSF) after the applied smearing is shown.}
\label{detectionmap}
\end{figure}

Figure~\ref{detectionmap} shows the relative flux map\footnote{Relative flux means excess events over background events. This quantity accounts for acceptance
differences between different parts of the camera} above an energy threshold of 150\,GeV around the center of the observations. The angular resolution of MAGIC 
for this analysis is 0.085\degr defined as one sigma of a Gaussian distribution, see \cite{MAGICperformance} for details.
The map was smeared with a two-dimensional Gaussian kernel with a sigma equivalent to that of angular resolution\footnote{The PSF shown in all skymaps is the sum in quadrature of the instrumental angular resolution and the applied smearing.}. 
Contours represent isocurves of test statistics (TS) evaluated from the excess of gamma-like events over a background model. This test statistic is Li\&Ma significance, 
applied on a smoothed and modeled background estimation. Its null hypothesis distribution mostly resembles a Gaussian function, but in general can have a somewhat 
different shape or width. The signal region is defined within 0.265\degr radius around the \textit{Fermi}/LAT position. This radius is selected in order to include 
the emission observed in the relative flux map. We compute an excess of $1371.7 \pm 122.5$ events inside the signal region, yielding a statistical significance 
of 11.4 standard deviations.
The centroid of the emission (black dot in Fig.~\ref{detectionmap}, statistical errors are represented by the ellipse) has been derived by fitting a $2$ dimensional Gaussian function to the map, prior to the smearing. As the centroid we find:
\[\mathrm{RA}=19.382 \pm 0.001 \,\mathrm{h} \,\,\,\,\,\,\,\,\, \mathrm{DEC}=14.191 \pm 0.015\, ^{\circ}\]
This deviates by 0.04\degr from the position reported by \textit{Fermi}/LAT, marked as the center of the sky map (green cross) (see Fig.~\ref{detectionmap}).

\begin{figure}
   \centering
   \includegraphics[angle=0,width=88mm]{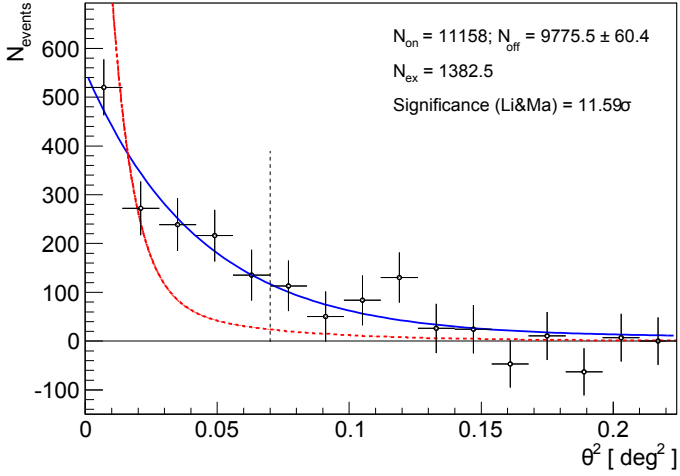}
   \caption{$\theta^2$ distribution of the excess events towards the centroid of the emission determined from figure~\ref{detectionmap}, showing a clear and extended signal.
The excess has been fitted by an exponential (blue curve)
to determine the extension. For comparison the shape of a point-like source with the same excess
determined from Monte-Carlo simulations is shown (red curve). The energy threshold of this analysis is
$150$\,GeV.}
\label{extension}
\end{figure}

To determine the extension of the source we computed the distribution of the squared angular distance $\theta^2$ between the arrival direction of the gamma-like events and
the centroid of the MAGIC source (see Fig.\ref{extension}), both for the integration area represented in Fig.~\ref{detectionmap}
and for a combination of signal-free regions from where we estimate the background. 

We then fit the difference between ON and OFF $\theta^2$ distributions using
an exponential function (corresponding to a Gaussian-shaped source).
For illustration, the shape of a point source with the same excess was calculated from Monte-Carlo simulations and is shown as comparison (red curve)
to the fit to the data (blue curve). After correcting for the angular resolution (0.085 degrees $>150$\,GeV) of the instrument the intrinsic extension of the source is
determined to be: $0.12 \pm 0.02_{\mathrm{stat}} \pm 0.02_{\mathrm{syst}}$ degrees.

\subsection{Spectrum}
\begin{figure}
   \centering
   \includegraphics[angle=0,width=88mm]{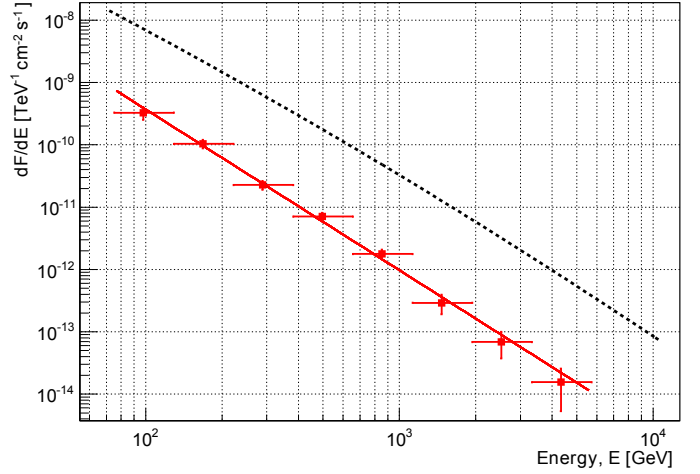}
   \caption{Differential energy spectrum of W51 obtained by MAGIC. The red points represent the differential flux points after unfolding.
The red line represents a power law fit to the data. The error bars represent the statistical errors. For comparison, the dotted line represents the spectrum of the Crab Nebula as shown in \cite{MAGICperformance}.}
\label{spectrum}
\end{figure}

We extracted the energy spectrum of the gamma-ray emission. The effective area was estimated using a
Monte-Carlo data set with photons simulated uniformly on a ring of 0.15 to 0.55\degr distance to the camera center. This accounts for
variations of the acceptance across the area of the source. The effect of using this ring Monte-Carlo compared to standard point-like ones turns out to lie well within
the statistical uncertainties. The spectrum needs to be unfolded in order to take into account the finite energy resolution and the energy bias of the instrument \citep{unfolding}.
The spectrum shown in Fig.~\ref{spectrum} starts at $75$\,GeV and is well described ($\mathrm{\chi^2}/\mathrm{NDF}= 5.26/6$) by a simple power law of the form:
\begin{equation}
 \frac{dN}{dE}=N_{0}\left(\frac{E}{1\,\mbox{TeV}}\right)^{-\Gamma}
\end{equation}
with a photon index of $\Gamma=2.58 \pm 0.07_{\tiny\mbox{stat}} \pm 0.22_{\tiny\mbox{syst}}$, and a normalization constant at 1\,TeV of
$N_{\tiny 0}=(9.7\pm 1.0_{\tiny\mbox{stat}})\times10^{-13}\mbox{cm}^{-2}\mbox{s}^{-1}  \mbox{TeV}^{-1}$.
This is the first time that the differential energy spectrum at VHE is published.
The energy threshold of MAGIC allows us to almost connect the spectrum to the \textit{Fermi}/LAT points \citep{fermi}.
The systematic error on the flux normalization is  15\%, which includes the systematic uncertainties of the effective area (11\%) and the background calculation.
In addition, the systematic uncertainty in the energy scale is estimated to be 17 \% at low ($\sim 100$\,GeV) and 15 \% at medium ($\sim 250$\,GeV) energies.
The integrated flux above 1\,TeV is
equivalent to $\sim3\%$ of the flux of the Crab Nebula above the same energy, and therefore agrees with the previous flux measurement by the H.E.S.S collaboration
\citep{W51Hess}. The spectral index measured by MAGIC agrees well with the one measured by \textit{Fermi}/LAT above 10 GeV \citep{david} of
$\Gamma=2.50 \pm 0.18_{\tiny\mbox{stat}}$.
The emission from W51 can be described by a single power law between 10\,GeV and 5.5\,TeV.

\subsection{Detailed morphology}
MAGIC reaches its best sensitivity in the energy range from $\sim$300 to $\sim$1000\,GeV. At energies of 300\,GeV the angular resolution of MAGIC is 0.075 \degr and it improves until reaching the saturation value of 0.054 \degr at energies above 1\,TeV.
We investigate sky maps in two energy ranges. The first map covers the estimated energy range from 300 to 1000\,GeV, and the second the energies above 1000\,GeV.
Both maps were smeared with a Gaussian kernel of a width equal to the angular resolution of the instrument in each energy range.

\begin{figure*}
   \centering
\includegraphics[angle=0,width=176mm]{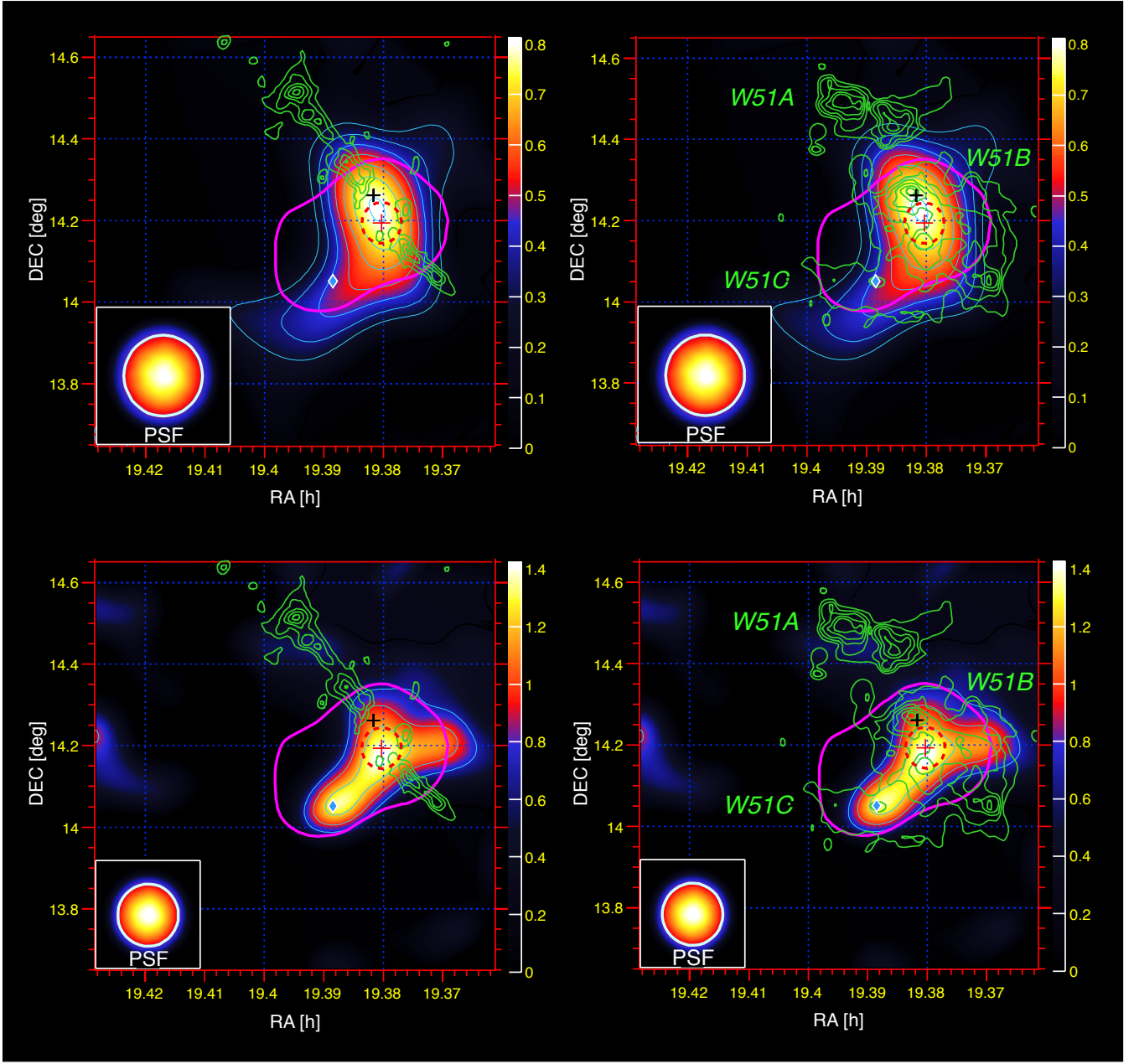}

  \caption{Relative flux maps: From $300$\,GeV to $1000$\,GeV (top) and $>1000$\,GeV (bottom).
On the left hand side the MAGIC data are combined with the $^{13}$CO (J=1-0) intensity maps from the Galactic Ring Survey 
(see http://www.bu.edu/galacticring/new\_index.html)
integrated between 63 and 72\,km$\mathrm{s}^{-1}$ shown as green countours.
On the right hand side the green contours represent the 21\,cm radio continuum
emission is shown from \citep{kooHI}.
In all maps the blue diamond represents the position of CXO J192318.5+140305 and the black cross the position of the OH maser emission \citep{kooChandra, maser}.
The red dashed ellipse represents the region of shocked atomic and molecular gas \citep{kooH2, kooHI}.
The 3 counts contour above 1\,GeV determined by \textit{Fermi}/LAT is displayed by the pink contour.
In each picture the gaussian sigma of a point-like source (PSF) after the applied smearing is shown.
The color scale (blue to red) represents the relative flux as measured with MAGIC.
In addition the TS contours (cyan) are shown starting at 3 and increasing by one per contour.}
\label{diffmaps}
\end{figure*}

In Fig.~\ref{diffmaps} (top panels) the relative flux map between 300 and 1000 GeV is shown. The overall shape of the emission appears to be elongated showing a
tail towards the lower left. The maximum of the emission coincides with the shocked-gas region, represented by the red dashed circle, where the lack of molecular material at the 
systemic velocity is clear (top left panel). The determined centroid and extension agree within statistical errors with those found above $150$\,GeV.

Above $1000$\,GeV (Fig.~\ref{diffmaps}, bottom panels) the centroid and the extension of the emission are in agreement with those obtained at lower energies. The
South-Eastern tail of the source, evident in the 300 to 1000\,GeV map, becomes a prominent feature coincident with the possible PWN CXO J192318.5+140305 at energies
above 1 TeV. However, the main part of the emission is still coincident with the shocked gas region. \\

While the centroid of the emission is consistent with the position of the shocked gas, we see a tail towards the PWN candidate. We note that, in any case, the VHE emission does not strictly follow the SNR shell
(as seen from the 21\,cm continuum emission represented by green contours in the right panels), nor does it follow the molecular
gas with the velocity expected due to Galactic rotation, as traced by the $^{13}$CO (green contours, left panels).
The tail seen towards the PWN rises the question of a possible substructure in the emission.

\subsubsection{Projections}
\begin{figure*}
\includegraphics[width=17.6cm]{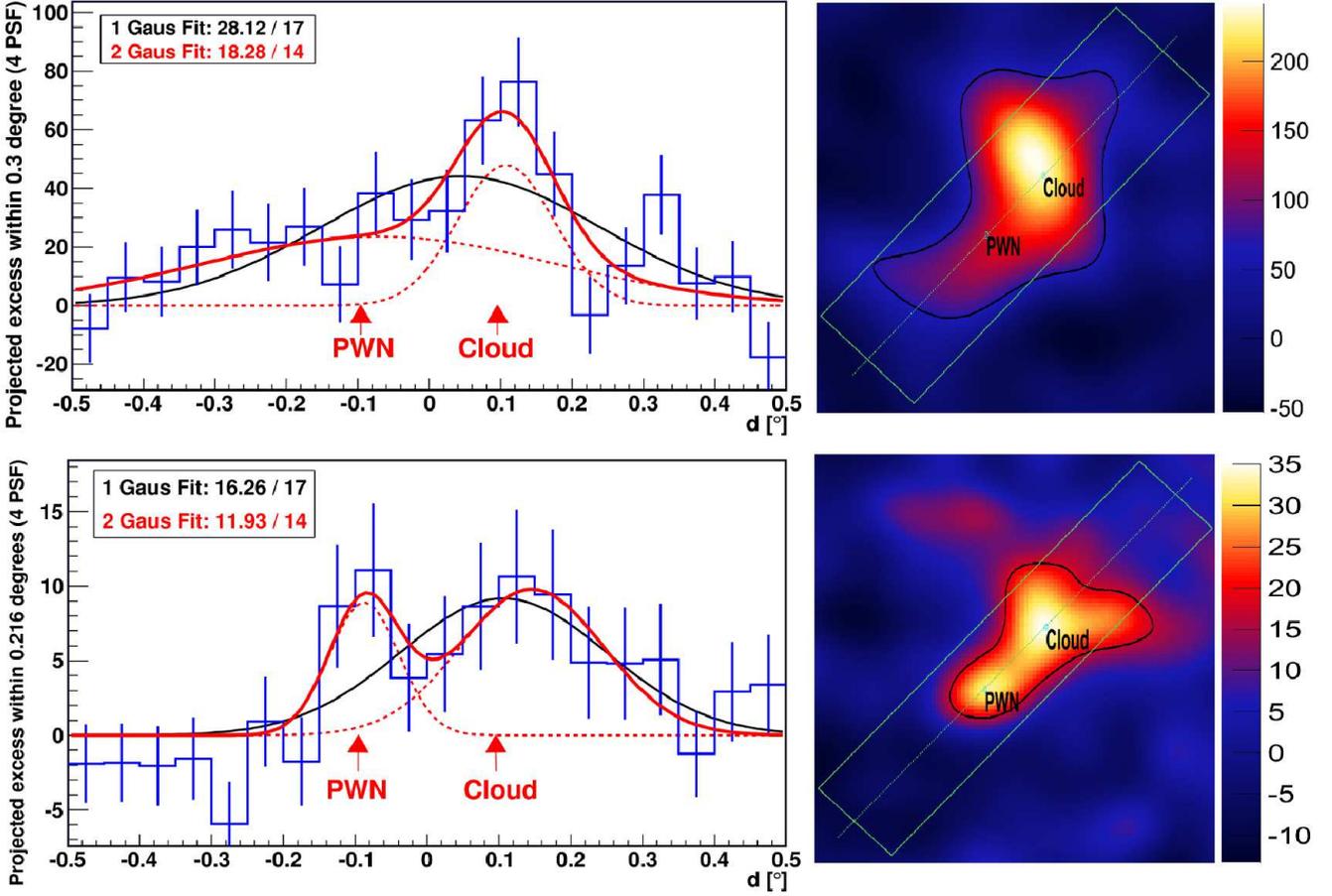}
  \caption{Projection of the excess inside the marked box in both differential sky maps:  $300$\,GeV to $1000$\,GeV (top) and above $1000$\,GeV (bottom) along the line 
connecting the PWN and the shocked-gas region described in \cite{kooHI,kooH2}.
 The projection is done with the unsmeared distribution.
The excess is fitted with one (black) and two (red) Gaussian curves. The positions of the shocked gas and the PWN are marked with red arrows.
On the right-hand side a sketch of the skymaps in both energy ranges is shown to illustrate the projected areas,
 as well as the position of the cloud and the PWN, respectively.
The box has a length of 1\degr and a width of 4 gaussian sigma of the instrumental PSF.
The sky maps show the smeared excess (for comparison with Fig.~\ref{diffmaps})
 with the black contour representing the 3 TS contour.
 }
\label{projectionmaps}
\end{figure*}

In order to investigate the source for underlying structures, we project the unsmeared excess distribution of the source along a line.
The line is 2\degr long divided in 40 bins with 0.05\degr width.
The orientation of the line is defined by the position of the PWN candidate and the centroid of the shocked clouds identified by
Koo \& Moon ($\mathrm{RA}=19.380 \,\mathrm{h}, \mathrm{DEC}=14.19^{\circ}$). Events within a distance of 2 gaussian sigma of the instrumental PSF to the line were projected.
Since the angular resolution is energy dependent, the width of the projected rectangle is 0.3\degr and 0.216\degr for the energy ranges from 300 to 1000\,GeV and above 1000\,GeV, respectively. OFF events were estimated from the background model.
The number of projected excess events is not the same as in the spectral calculation, where we used a circular region of 0.265\degr radius around the center of the
observations. Therefore, the projected excess does not allow for direct determination of the fluxes from specific regions of the map.
The projection has been carried out in both
energy ranges independently on the unsmeared excess distribution and is shown in Fig.~\ref{projectionmaps}.

We fit the projection alternatively using one and two Gaussian functions. $\chi^{2}/\mathrm{d.o.f.}$ values are 28/17 (one Gaussian) and 18/14 (two Gaussians) for the medium-energy
range and 16/17 (one Gaussian) versus 12/14 (two Gaussians) for the high-energy events.
The data are very well described with the two Gaussian functions, where the centroid of the individual functions coincides within statistical errors
with the position of the shocked gas and the
PWN. The tail-like feature towards the possible PWN is more peaked in the energy range above 1000\,GeV.

The statistics are not sufficient to clearly discriminate between an extended source of Gaussian excess, an extended source of a more complicated shape, or two
 individual sources. However, the fact that there is no region of dense gas close to the PWN makes it difficult to explain the enhancement of TeV emission in this area
under the assumption of uniform CR density. A possible scenario of two emission regions could manifest in different spectral behaviours.

\subsubsection{Energy spectra of individual regions}
To quantify the results obtained from the projections we investigated in more detail the spectral properties of the detected signal, we concentrated on two
individual regions within the source and analyzed them separately.
One was defined to cover the shocked cloud region with centroid at $\mathrm{RA}=19.380 \,\mathrm{h}, \mathrm{DEC}=14.19^{\circ}$;
this will be called the \textit{cloud} region. The second one was defined by the position of CXO J192318.5+140305 and will be called the \textit{PWN} region.
To avoid contamination from the surrounding emission, and their possible spread due to the worse angular resolution at lower energies, we use an integration radius of
0.1\degr. We compared the same analysis on data of the Crab Nebula and find that such a region contains at least 70\% of the excess from a point-like source above 300\,GeV.
For an easier comparison, the integration radii were chosen
 to be the same.  The distance between the chosen positions is 0.19\degr. There is an area of overlap of $1.7\%$ compared with the integration area of each region, therefore they can be treated as independent.
The combined areas of both regions represent about 57\% of the area used to determine the overall spectrum.

The small distance between the regions and a very similar average distance to the camera center
allow us to assume the same acceptance of gamma-like events for both regions, at least within 5\%.

For each individual region we determined the amount of excess events above three different energies, and calculated the contribution to the overall emission.
The resulting values are shown in Table~\ref{ratios}. Excesses used to calculate these ratios show a significance of at least $2.9\sigma$.

\begin{table}[H]
\caption[]{Number of excess events determined for the \textit{PWN}-region and the \textit{cloud}-region and their contribution to the overall emission.
Within the statistical errors we do not detect a significant energy dependence on their contributions to the overall excess.}
         \label{ratios}
\centering{}
\begin{tabular}{lllll}
\hline\hline
$\mathrm{\textit{E}\,[GeV]}$  & \textit{cloud} & \textit{PWN} & \textit{cloud}/all [\%] & \textit{PWN}/all [\%]\\
\hline
$>300$     &  $200 \pm 30$ & $132 \pm 25$ &  $30 \pm 5$ & $19 \pm 4$  \\
$>500$  &  $116 \pm 17$ & $79 \pm 17$  &  $32 \pm 6$ & $22 \pm 5$  \\
$>1000$  & $48 \pm 10$ & $27 \pm 10$  & $43 \pm 12$ & $24 \pm 10$   \\
\hline
\end{tabular}
\end{table}

The excess contribution arising from the cloud region is about 30\% and shows no dependence on energy.
We performed a spectral analysis of a point source for the cloud region above 350 GeV.
The emission can be well described by a pure power law with a flux normalization constant at 1\,TeV of
$N_{\tiny\mbox{cloud}}=(4.3\pm 0.9_{\tiny\mbox{stat}})\times10^{-13}\mbox{cm}^{-2}\mbox{s}^{-1}  \mbox{TeV}^{-1}$. The integrated flux above 350 GeV is equivalent to 1.2\%
 of the flux of the Crab Nebula.
The spectral index of the cloud emission is $-2.4\pm0.5_{\tiny\mbox{stat}}$ and agrees within statistical uncertainties with the spectral index of the overall emission.

Assuming a point-like emission,
the flux from the \textit{PWN} region above 350\,GeV is equivalent to 0.7\% of the flux of the Crab Nebula representing about 20\% of the overall observed emission. The emission between 350\,GeV and 2\,TeV can be well described by one single power
law with a spectral index of $-2.5\pm0.6_{\tiny\mbox{stat}}$ and a flux normalization at 1\,TeV of
$N_{\tiny{PWN}}=(2.3\pm 0.8_{\tiny\mbox{stat}})\times10^{-13}\mbox{cm}^{-2}\mbox{s}^{-1}  \mbox{TeV}^{-1}$.

The excess contribution of each of the regions shows no dependence on energy, suggesting no intrinsic morphological changes in the energy ranges investigated here. This is in
agreement with the spectra, with the differential maps, and with the projections of the excess distribution.
We want note that the number of excess events
within the \textit{PWN} region and the \textit{cloud} region (Table~\ref{ratios}) agrees within statistical errors with the projected
excess (Fig.~\ref{projectionmaps}) found within $\pm0.1$ degree from the PWN and the cloud positions, respectively.
By looking at the skymaps (Fig.~\ref{projectionmaps}) only, the emission around the PWN seems to be more intense above 1\,TeV.
However this can be explained by the worse angular resolution at lower energies and by a much higher signal-to-noise ratio at the higher energies.

\section{Discussion}
\label{sec:Model}

Before modeling the multi-wavelength emission we address shortly the possibility
that the PWN alone is the source of all the emission. Then we qualitatively discuss the possibility of a contribution of the PWN to the overall emission
and justify the approach using a one-zone model (i.e. one homogeneous emission region filled with one particle distribution per
species) to investigate the processes underlying the emission.

First, we want to recall that we have found no spectral or morphological energy dependence. In order to assess whether the VHE emission can be originated only by the PWN candidate, we consider the estimated rate of rotational energy loss
\textit{\.{E}}$=1.5\times10^{36}$\,erg\,s$^{-1}$ estimated by \cite{kooChandra} with the empirical relation from \cite{seward}. Given the
observed luminosity of the order of $\sim10^{36}$\,erg\,s$^{-1}$ reported in \cite{fermi}, it seems unlikely that
the PWN alone is the source of all gamma-ray emission since, it would require an
extremely high efficiency in the conversion from rotational energy into gamma rays.

Second, we consider the possibility of a two-zone model. The \textit{PWN} region can account for the 20\% of the gamma-ray emission confined
in a point-like source, however the brightest part is the \textit{cloud} region. This scenario would require an \textit{\.{E}} conversion
into gamma rays of the order of 10\%, in agreement with the generally accepted value.

 With the current statistics and resolution, it cannot be established if there is a spectral
difference between the \textit{cloud} and the \textit{PWN} region, but the contribution of the \textit{PWN} is in any case small.
For the reasons above, the simplest approach is to assume one overall particle distribution underlying the emission we observe.
This assumption introduces an error in the
flux normalization of about 20\% in case part of the emission originates from the PWN candidate; this uncertainty lies within the statistical and
systematic errors of the MAGIC measurement.

\subsection{Model description}
\label{sec:ModelDescription}

We model the SNR as a sphere homogeneously filled with hydrogen, helium and electrons, with respective average number densities
${n}_{\text{H}}$, ${n}_{\text{He}}$ and ${n}_{e^-}$. For the relative abundances of helium we assume the cosmic abundance ratio
${n}_{\text{He}} = 0.1 \: {n}_{\text{H}}$. For the electron ratio we assume full ionization of the medium, such that ${n}_{e^-} = 1.2 \: {n}_{\text{H}}$.
The magnetic field $B$ is assumed to be homogeneous inside the sphere; \cite{W51Bfield} derived an upper limit for it of $B_{\parallel} <$ 150 $\mu$G, but
\cite{W51MaserBfield} measured a local magnetic field as high as 1.5--1.9 mG towards the maser sites.

The geometric model of the SNR and molecular cloud interaction region as proposed by \cite{kooHI}, describes a scenario in which the spherical blast wave of the supernova
explosion interacts with part of the cylindrical molecular cloud contained inside the SNR volume.
\cite{W51GiantMC} estimated the total mass of the molecular cloud to be $m_{\text{cloud}} = 1.9 \times 10^5 \: \text{M}_{\odot}$.
From the radio measurements in \cite{W51Radio} the angular extent
of the partial radio shell of the SNR is known to be $\theta \approx 30^\prime$ .
We see a clear displacement between the morphology presented here and the center of the spherical extended SNR as seen in thermal X-ray emission \citep{W51Rosat}.
The maximum of the emission is located at the interaction region of remnant and the molecular cloud.
Therefore we conclude that the size of the remnant is not physically related to the size of the VHE emission region.
We adopt the intrisnic extension determined in this work to determine the radius of a spherical emission zone.
Assuming a distance to W51C of $5.5$ kpc, as measured by
\cite{parallax} and \cite{W51SpectDistance}, the radius of the sphere is estimated to be $ 24 \: \text{pc}$.

The explosion energy of the
SNR has been estimated in \cite{W51Rosat} as $E_{\text{SN}}\approx 3.6 \times 10^{51} \: \text{erg}$, using both a Sedov and an evaporative model to derive the parameters of the SNR.
We will compare this value with the one obtained from the integral of our initial spectra, after we fix the normalisation constants; we will determine how much of the initial explosion energy of the supernova has been converted into particles ($W_e$, $W_p$).
The different parameters of the supernova, of the SNR, and of the molecular cloud are summarized in Table~\ref{tab:W51Parameters}.

\begin{table}[H]
\centering
\caption{Parameters of the W51C supernova, supernova remnant and molecular cloud.}
\begin{tabular}{lcr}
\hline\hline
Parameter			&	Value				&	Reference\\
\hline
age					&	$\approx$ 30 000 yr	&	\cite{W51Rosat}\\
$E_{\text{SN}}$		&	$\approx 3.6 \times 10^{51} \: \text{erg}$	&	\cite{W51Rosat}\\
$d$					&	5.5 kpc				&	\cite{parallax}\\
 					& 						&	\cite{W51SpectDistance}\\
$\theta$ (radio)	&	$\approx 30^\prime$		&	\cite{W51Radio}\\
$B_{\parallel}$		&	$<$ 150 $\mu$G		&	\cite{W51Bfield}\\
$B$ (at masers)		&	1.5--1.9 mG			&	\cite{W51MaserBfield}\\
$\alpha_{r}$		&	$\approx -0.26$ 	&	\cite{W51Radio}\\
$m_{\text{cloud}}$	&	$1.9 \times 10^5 \: \text{M}_{\odot}$		&	\cite{W51GiantMC}\\
\hline
\end{tabular}
\label{tab:W51Parameters}
\end{table}

We model the spectral  energy distribution folding input spectra of accelerated particles with cross sections of processes yielding photons; this includes synchrotron radiation, inverse Compton scattering (IC), non-thermal bremsstrahlung and $\pi^0$ decay \citep{BlumenthalGould1970, BaringReynolds1999, KelnerAharonian2006}.

For IC, we consider three seed photon fields: the cosmic microwave background ($kT_{\text{CMB}} = 2.3 \times 10^{-4}$ eV, $u_{\text{CMB}} = 0.26 \: \text{eV} \: \text{cm}^{-3}$), infrared
($kT_{\text{IR}} = 3 \times 10^{-3} \: \text{eV}, u_{\text{IR}} = 0.90 \: \text{eV} \: \text{cm}^{-3}$) and optical ($kT_{\text{OPT}} = 0.25 \: \text{eV},
u_{\text{OPT}} = 0.84 \: \text{eV} \: \text{cm}^{-3}$), with temperatures and energy densities for the infrared and optical components adopted from \cite{fermi}.
Bremsstrahlung is computed on a target of electrons and ions.
For the $\pi^0$ production cross section, we use the parametrization of Kelner \& Aharonian (2006) with a constant nuclear enhancement factor of 1.85 \citep{Mori2009}.

The multi-wavelength data considered here include radio continuum measurements \citep{W51Radio}, high-energy observations by the \textit{Fermi}/LAT \citep{fermi} and
the new VHE data taken with MAGIC, presented in this paper. Included is also one data point by MILAGRO \citep{MILAGRO}.
Note that the lowest energy radio data point may be affected by free-free absorption, see \cite{W51Radio} or \cite{W51RadioCopetti1991}, which we do not consider here. 
However, this single point does not affect the fitting of the radio data. The radio measurements in \cite{W51Radio} indicate a spectral index of
$\alpha_{r} \approx -0.26$ (as defined by $S_{\nu} \propto \nu^{\alpha_r}$). This can be attributed to electrons emitting synchrotron radiation and fixes the initial
 power-law index of the electron spectrum to $s \approx 1.5$. We adopt this value both for electrons and protons.

As an upper limit for the non-thermal X-ray emission we consider the integrated thermal X-ray flux of the whole remnant as measured by \textit{ROSAT} \citep{W51Rosat}
converted into a differential flux in the sub-keV range. We use the thermal emission observed by Chandra from CXOJ192318.5+140305 as an upper limit to
the non-thermal emission of the possible PWN. The MILAGRO measurement has a significance of 3.4 $\sigma$, was derived assuming a
gamma-spectrum $\propto E^{-2.6}$ without a cut-off and is given at an energy of 35 TeV. For details see \cite{MILAGRO}.

We consider seperate scenarios in which one of the following emission processe dominates over the others,
pion decay, inverse Compton, or Bremsstrahlung. The models discussed here are obtained using as equilibrium particle spectra a broken power law
with an exponential cut-off, both for electrons and protons, of the form:

\begin{equation}
	\frac{d N_{e,p}}{d E_{e,p}} = K_{e,p} \left( \frac{E_{e,p}}{E_0} \right)^{-s} \left[ 1 + \left( \frac{E_{e,p}}{E_{\text{br}}} \right)^{\Delta s} \right]^{-1} \exp \left[ - \left( \frac{E_{e,p}}{E_{{\text{cut},{\it e,p}}}} \right) \right]
	\label{eq:BrokenPowerLaw}
\end{equation}

The spectral index changes here from $s$ to $s + \Delta s$ at an energy $E_{\text{br}}$ with a smooth transition. The exponential cut-off at
$E_{\text{cut,{\it e,p}}}$ reflects the roll-off of the particle spectrum near the maximum energy, arising from the acceleration and confinement mechanism,
as well as energy losses.

The break energy $E_{\text{br}}$ is fixed from the \textit{Fermi}/LAT data, while the new MAGIC data allow us to fix the spectral
break $\Delta s$.
A spectral break in the particle spectrum at these energies is traditionally thought to be inconsistent with both standard or non-linear diffusive shock acceleration theory, see \cite{Malkov2001} and references therein.
However, \cite{Malkov2011} have recently proposed a mechanism which can also explain a spectral break in the cosmic ray spectrum of $\Delta s=1$ by strong
ion-neutral collisions in the surroundings of a SNR, leading to a weakening in the confinement of the accelerated particles. The spectral break that we
 have derived here is $\Delta s$ = 1.2, not far off this prediction, giving a hint that this mechanism might be responsible for the observed break.
Note also that other authors have proposed scenarios in which the CR spectrum, and consequently the gamma-ray spectrum, can show one or more spectral breaks,
for example due to finite-size acceleration or emission region \citep{ohira} or energy dependent diffusion of run-away CRs from the remnant
\citep{gabici, Aharonian1996}.

The luminosity of W51C in the energy range 0.25 GeV -- 5.0 TeV, which is roughly the energy range of the \textit{Fermi} and MAGIC data,
is $L_{\gamma} \approx 1 \times 10^{36}$\,erg\,s$^{-1}$, assuming a distance of 5.5 kpc, which is one of the highest compared with other SNRs.

\subsection{Adjustment of model parameters}
\label{sec:ModelResults}

First we consider the case where the emission is dominated by leptonic emission mechanisms. We find the same problems already reported by \cite{fermi}, namely that it 
cannot reproduce the radio and gamma-ray data simultaneously. Furthermore these models
need an unusually high electron to proton ratio of the order of one.

\begin{figure}
 \centering

	\includegraphics[width=0.5\textwidth]{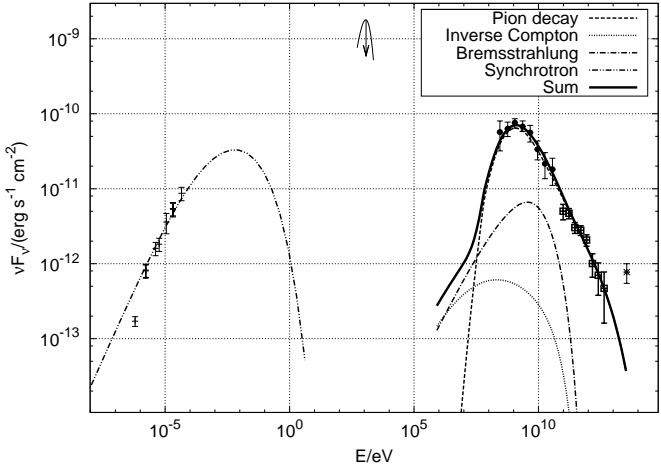}

 \caption{Model of the multi-wavelength SED in the hadronic-dominated scenario.
The dashes with error bars are 21cm radio continuum, circles represent \textit{Fermi}/LAT data, squares are the data obtained in this work and the star represent the MILAGRO data point.
The upper limit in the X-ray regime is obtained from \textit{ROSAT} data as discussed in the text. The details of the scenario are discussed in the text.}
 \label{fig:Fits}
\end{figure}

When we model the emission with pion decay as the dominant process, both radio and gamma-ray emission can be reasonably reproduced, as shown in Fig.~\ref{fig:Fits}.
A hadronic scenario is particularly interesting, as the shock-cloud interaction naturally favors a CR-matter interaction mechanism.
Moreover, the parameters used in this model, see Table~\ref{tab:W51Broken} are a reasonable description of the interstellar medium around W51. 

\begin{table}[H]
\centering
\caption{Parameters used in the modeling of the multi-wavelength spectral energy distribution for the hadronic scenario.
The power-law index before the break is $s$ = 1.5 for both protons and electrons. $E_0$ = 10 GeV.
The total kinetic energy of the particles was integrated for $E_{\text{kin}} >$ 100 MeV both for electrons and protons.}
\begin{tabular}{lc}
\hline\hline
Parameter			& Value\\
\hline
$K_e/K_p$                       & 1/80\\
$\Delta s$                      & 1.2\\
$E_{\text{br}}$ [GeV]            & 10\\
$E_{\text{cut},\it{e}}$ [TeV]     & 0.1\\
$E_{\text{cut},{\it p}}$ [TeV]    & 120\\
${B}$ [$\mu$G]                  & 53\\
${n}$ [cm$^{-3}$]               & 10.0\\
$W_e$ [$10^{50}$ erg]           & 0.069\\
$W_p$ [$10^{50}$ erg]           & 5.8\\
\hline
\end{tabular}
\label{tab:W51Broken}
\end{table}

Compared to the hadronic model suggested in the work of \cite{fermi}, the main difference is the index of the partcile distribution after the break.
The spectrum after the break is more precisely determined by the data presented here.
The index we obtain is harder, allowing for the explanation of all the gamma-ray data up to the end of the MAGIC spectrum.

\begin{figure}
 \centering
 \includegraphics[width=0.5\textwidth]{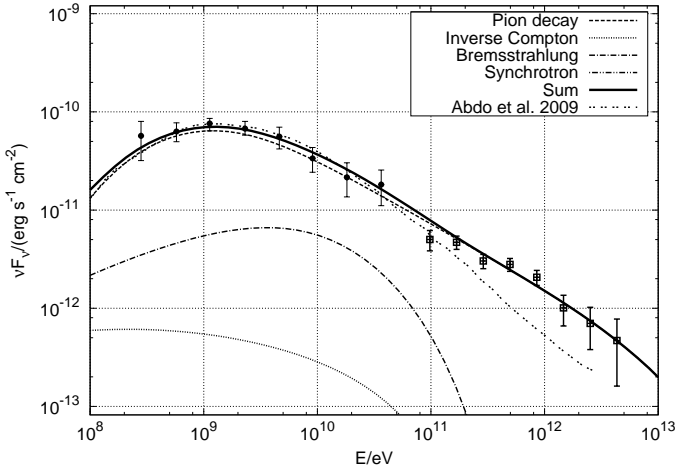}
 \caption{Detailed view of the hadronic model in the high energy and VHE region. For comparison to the hadronic model by \cite{fermi}, shown as double dotted line.
This models ends at the highest energy shown in that publication. The major difference between our model and that of \cite{fermi} is the harder particle spectrum
above $\sim$100\,GeV, which is now precisely constrained by the measurements presented here.}
 \label{fig:HadronicDetail}
\end{figure}

A detailed view of the high energy and VHE region is shown in Fig.~\ref{fig:HadronicDetail}. It shows that the index above the break is clearly determined by the data
presented here. In addition, the hadronic model by \cite{fermi} is displayed. In addition to the good aggrement between the model and the data, the plot shows that the results presented 
here clearly improve the determination of the underlying particle distribution. In this scenario a cut-off energy of $E_{\text{cut,{\it p}}} \geq$ 100 TeV is needed to fit the MAGIC data, indicating the existence of protons at least to this energy.

The precise cut-off energy of the electron spectrum $E_{\text{cut,\it{e}}}$ is not well constrained, since the synchrotron peak is not resolved.
Therefore, the energy $E_{\text{cut,\it{e}}}$ used here is only a lower limit, as enforced with the radio data.
However, a 1 TeV electron in a magnetic field of 50\,$\mu$G has a lifetime of about 4700 years, determined by synchrotron losses.
This value is much lower than the age of the remnant, suggesting that for such high energies the electron spectrum should develop a break, with the consequent
spectral steepening at higher energies. Assuming constant electron injection over time, the electron spectrum steepens at ~100 GeV by a factor $1/E$. 
This yields a very similar gamma-ray emission as in the hadronic model presented here, even for a higher value of $E_{\text{cut,\it{e}}}$.

\subsection{Physical outcome of the models}
\label{sec:Summary}
We discuss what general conclusions can be drawn from the model which fit the data: the hadronic scenario.

The volume-averaged hydrogen density is obtained as a parameter of the fit. From that, we compute the volume filling factor $f$, which is the
 fraction of the mass of the clumpy molecular cloud that is contained inside the SNR interaction volume (defined as volume of the emission zone) as
$f = {n}_{\text{H}} V \left( m_{\text{H}}^{\text{atom}} + 0.1 m_{\text{He}}^{\text{atom}} \right) / m_{\text{cloud}} \approx$  0.11. Here
$m_{\text{cloud}}$ is the total mass of the molecular cloud, $V$ is the volume of the radiation sphere and $m_{\text{H}}^{\text{atom}}$, $m_{\text{He}}^{\text{atom}}$ are the masses of a hydrogen and helium atom, respectively.
This would imply that around 11 \% of the mass of the molecular cloud is contained in the emission volume and is interacting with the SNR.
This value is consistent with the filling factors of around 8--20 \% for other SNRs interacting with molecular clouds, obtained by other  authors \citep{uchiyama}.

The total amount of kinetic energy in electrons and protons is about $16 \: \%$ of the explosion energy of the supernova. This fraction is just slightly higher than
the value normally assumed, of around $10 \: \%$, of the explosion energy converted into CRs to maintain the observed flux of Galactic CRs \citep{Hillas2005}.
The proton to electron ratio is not far from value observed at earth of $K_p/K_e \approx 50$, see for example \cite{Simpson}. 

Since the hadronic gamma-ray emission is proportional to the product of the kinetic energy in protons and the density of the medium, this parameters are striclty 
correlated. Assuming that the complete mass of the molecular cloud acts as target material ($f$=1), this would imply a density of ${n}$=100 cm$^{-3}$. 
Therefore the lower limit
of the energy in relativistic protons is about $1.6 \: \%$ of the explosion energy of the supernova. We note that such a scenario would need either a higher magnetic 
field (${B}$ $\sim$ 150 $\mu$G) or a much lower electron to proton ratio ($K_e/K_p \sim$ 1/800) to still reproduce the broadband emission. In addition, the morphology 
presented in this work shows that only a fraction of the molecular cloud is emitting VHE gamma emission (see Fig.~\ref{diffmaps}). 
Therefore we conclude that the amount of kinetic energy in protons is clearly above this lower limit and in the order of 10--20\%.

In the scenario investigated here all of the gamma-ray emission was attributed to $\pi^0$ decay. It was not possible to model the broad-band
 emission with a purely leptonic scenario. The radio data could not be fitted and the model parameters were not physically reasonable (too low density ${n}_{\text{H}}$,
 too high energy content $W_{e}$ in electrons, too low magnetic field ${B}$). However, that could also point to problems in the modeling, especially to
oversimplifications concerning the homogeneity of the medium and of the magnetic field.

We conclude that the \textit{Fermi}/LAT data and the MAGIC data can be explained in terms of hadronic interactions of high-energy protons with the molecular cloud and
subsequent decay of neutral pions. With the current data it is not possible to decide what process causes the hint of emission observed by MILAGRO which, if confirmed at 
this flux level, would require the introduction of an additional component at the highest energies.

\subsection{Discussion on the acceleration process}

Following the result of the modeling we assume the observed gamma-ray emission to be of hadronic origin.
As mentioned in Section~\ref{intro}, there are two main possible scenarios: a cloud illuminated by runaway CRs or acceleration of CRs
in the shock wave propagating through the cloud.

In the first case, CRs escaping the SNR will homogeneously fill a sphere with a radius ${R_{d}}\sim\sqrt{4{D\,t}}$ where $D$ is the diffusion coefficient and $t$ is the time since
particles are diffusing \citep{gabici_diff}. For a distance of $5.5\,\mathrm{kpc}$ and $10\, \mathrm{TeV}$ protons, responsible for gamma-ray emission of
$1\, \mathrm{TeV}$, the radius of this sphere would be about $350$\,pc, assuming the average Galactic CR diffusion coefficient at 10 TeV to be
$\sim 3\times 10^{29} \mathrm{cm}^2$ s. Here we assumed that the high-energy particles escape the SNR early enough such that the diffusion time can be approximated
to be the age of the SNR. The distance between the maximum of the emission measured by MAGIC above $1\,\mathrm{TeV}$ and the assumed center of the SNR
(RA=19.384 h, DEC=14.11\degr) is about $8\,\mathrm{pc}$. The distance to other parts of the SNR/cloud complex W51C/B is of similar order.
This implies that the complete cloud should be uniformly illuminated by CRs. As can be seen in Fig.~\ref{diffmaps} we do not detect the complete W51B/C complex
at energies above 1 TeV (lower right skymap): parts of the outer regions, both on the side towards the SNR and on
the opposite side, do not emit gamma radiation.
In the scenario of runaway CRs, we would expect diffusion from the SNR to W51A (northern region in the 21cm emission); no significant emission from W51A is detected.
However, the distance of the regions A and B is measured with an error of the order of hundreds of parsecs, which means that
the relative distance between the two could be high enough to explain the lack of diffusion from one to the other.
The scenario of runaway CRs also can not explain the incomplete illumination of W51B/C, especially towards the outer regions.

Concerning the acceleration of CRs in the shocked cloud scenario, the gamma radiation should be originated very close to the acceleration site of the radiating particles due to
 the high density of the surrounding medium. This is in agreement with the morphology described in this work. The unusually high ionization reported by \cite{crionization}
close to the maximum VHE emission region indicates the presence of freshly accelerated low-energy protons. The missing
emission towards the edges of the cloud could be explained with a lower diffusion coefficient in the shocked cloud region, or with a shielding effect, either of which is
possible in a surrounding medium of high density.

Both the morphology at TeV energies and the measured high ionization
are hints for an ongoing acceleration. This suggests that the particle distribution, whose gamma emission we observe, may represent the source spectrum
of cosmic rays currently being produced in W51. However, the differentiation between ongoing acceleration of particles
 in the shocked region or reacceleration of already existing CRs, like in the \textit{crushed cloud} scenario \citep{uchiyama}, in the same region is not obvious and is not addressed in this work.

\section{Conclusions}

MAGIC has performed a deep observation of a complex Galactic field containing the star-forming regions W51A and W51B, the SNR W51C and the possible PWN CXO J192318.5+140305.
As a result of this observation, emission of gamma rays above 150\,GeV has been detected with 11 $\sigma$ statistical significance.
The spectrum of this emission has been measured between 75\,GeV and 4\,TeV. Spectral points are well fitted with a power law with a photon index of 2.6, compatible with the
\textit{Fermi}/LAT measurement between 2 and 40\,GeV. The spectrum measured by MAGIC allows for the first time a precise determination of the spectral slope of the
underlying particle distribution above the spectral break measured at around a few GeV by \textit{Fermi}/LAT.

The MAGIC source spatially coincides with those previously reported by H.E.S.S. and \textit{Fermi}/LAT. We are able to restrict
the emission region to the zone where W51C interacts with W51B and, in particular, to the region where shocked gas is observed. This clearly pinpoints the origin of the
emission to the interaction between the remnant and the molecular cloud.

Non-thermal X-ray emission which could help to trace the relativistic electron distribution was found only from a compact region around the position of the possible
PWN CXOJ192318.5+140305 \citep{kooChandra}. The MAGIC source exhibits a morphological feature extending towards CXO J192318.5+140305, more prominent in the image at higher
energies.

The projection of the gamma-like
events on the line connecting the putative PWN and the centroid of the shocked clouds shows a hint of an underlying distribution that may be described as the sum of two Gaussian functions.
However, the existence of two independent, resolved sources cannot be statistically established. We thus investigate the contribution to the total excess of two regions of
0.1\degr radius centered on the \textit{cloud} region and the \textit{PWN} region. We find that they contribute about 30\% and 20\% of the total emission, respectively, and the
contribution is not energy dependent within the uncertainties. Spectra of the individual regions above 350 GeV could be obtained, but do not allow for detailed
conclusions due to the weak individual fluxes.
Given the small possible contribution of the PWN candidate in the energies investigated in this work, it is very unlikely that the main conclusion drawn here will be
significantly affected even if the PWN contribution can be established.

MAGIC observations determine the VHE spectral energy distribution of W51 over more than one order of magnitude in energy. We have produced a physically plausible model of the emission of
the SNR by considering a spherical geometry and uniform distribution of the ambient material. We note that this system is clearly anisotropic (as seen in the
 multi-wavelength data), and more detailed modeling may achieve a better description of the source.
We find that the VHE emission from W51C cannot be explained by any of the considered leptonic models.
The emission is best described when neutral pion decay is the dominant gamma-ray production mechanism. In the proposed model,
the SNR has converted about 16\% of the explosion energy into kinetic energy for proton acceleration and the emission zone engulfs a 10\% of a molecular cloud of
$10^{5}$ solar masses, which provides the target material.
In this scenario, protons are required to reach at least an energy of the order of 100 TeV to produce the observed emission.

The morphology of the source cannot be explained by CRs diffusing from the SNR to the cloud.
It can instead be qualitatively explained with VHE gamma-ray emission being produced
 at the acceleration site of CRs. This involves ongoing acceleration of CRs or re-acceleration of already existing CRs at the shocked cloud region.
Given the high luminosity of this source and its plausible hadronic origin, we conclude that W51C is a prime candidate cosmic ray source in the Galaxy.

Finally, we want to give a short outlook and address the a few issues connected with W51C.
The detection of neutrinos from this source would be the final proof about the hadronic nature of the emission. But, according to the calculations by
\cite{neutrinos}, the chances for detection are low.
However, also an extension of the high-energy gamma emission towards lower energies, as performed for example in \cite{martina}, may also provide more clues to the nature
of particle acceleration in this region.
To reveal the morphology and the possible emission of the PWN, more data at energies above 1 TeV are necessary.
Extension of the spectrum towards higher energies would constrain the maximum achievable energy in the system and might shed light on the meaning of the MILAGRO
measurement, which cannot be accommodated in the theoretical framework proposed here.

\begin{acknowledgements}
We would like to thank the anonymous referee as well as the Associate Editor M. Walmsley for fruitful comments and suggestions.
We would like to thank the Instituto de Astrof\'{\i}sica de
Canarias for the excellent working conditions at the
Observatorio del Roque de los Muchachos in La Palma.
The support of the German BMBF and MPG, the Italian INFN,
the Swiss National Fund SNF, and the Spanish MICINN is
gratefully acknowledged. This work was also supported by
the Marie Curie program, by the CPAN CSD2007-00042 and MultiDark
CSD2009-00064 projects of the Spanish Consolider-Ingenio 2010
programme, by grant DO02-353 of the Bulgarian NSF, by grant 127740 of
the Academy of Finland, by the YIP of the Helmholtz Gemeinschaft,
by the DFG Cluster of Excellence ``Origin and Structure of the
Universe'', by the DFG Collaborative Research Centers SFB823/C4 and SFB876/C3,
and by the Polish MNiSzW grant 745/N-HESS-MAGIC/2010/0.

This publication makes use of molecular line data from the Boston University-FCRAO Galactic Ring Survey (GRS).
The GRS is a joint project of Boston University and Five College Radio Astronomy Observatory, funded by the National Science Foundation under grants
 AST-9800334, AST-0098562, AST-0100793, AST-0228993, \& AST-0507657.
\end{acknowledgements}

\bibliography{W51_final.bib}
\bibliographystyle{aa}

\end{document}